\documentclass[11pt]{article}
 
\usepackage{epsfig}
\usepackage{amsmath}
\usepackage{amssymb}
 
\newcommand{\one}{1\!\!1}

\begin{document}


\renewcommand{\thefootnote}{\fnsymbol{footnote}}

\title{ Composite boson dominance in relativistic field theories} 

\author{ Sergio Caracciolo\\
\small\it  Universit\`a degli Studi di Milano - Dip.~di Fisica and INFN,\\ 
\small\it  via Celoria 16, I-20133 Milano, Italy\\ 
\small\tt Sergio.Caracciolo@mi.infn.it      \\ 
{\protect\makebox[5in]{\quad}}
\and 
 Victor Laliena\\ 
\small\it  Universidad de Zaragoza - Depto. de Fisica, \\
\small\it  Pedro Cerbuna 12, 50009 Zaragoza, Spain \\ 
\small\tt laliena@unizar.es  \\    
{\protect\makebox[5in]{\quad}}
\and
Fabrizio Palumbo\\
\small\it  INFN, Laboratori Nazioni Frascati, \\ 
\small\it  P.~O.~Box 13, I-00044 Frascati, Italy \\ 
\small\tt Fabrizio.Palumbo@lnf.infn.it    
}

\date{\today}
\maketitle

\begin{abstract}
\noindent
We apply a new bosonization technique to  relativistic field theories of fermions whose partition 
function is dominated by bosonic composites, and derive the effective action for these  bosons. 
The derivation respects all symmetries, including gauge invariance, with the exception of
Euclidean invariance which must be checked a posteriori. We use a lattice regularization which
should make  applications to gauge theories easier.
We test the method  on a fermion field theory with quartic interaction in the limit of large number of flavours $N_f$, and show that it 
reproduces the exact results in the bosonic sector, namely condensation of a composite
boson with the right mass which breaks the discrete chiral invariance of the model. Moreover we 
determine the structure function of the condensed composite, whose spatial part
turns out to be identical to that of the Cooper pairs of the BCS model of superconductivity. 
\end{abstract}

\vfill\eject

\renewcommand{\thefootnote}{\arabic{footnote}}
\setcounter{footnote}{0}\textit{}

\section{Introduction}

 There are many theories involving fermions whose low energy excitations are 
dominated by bosonic
modes. Historically the first case of great general relevance is superconductivity in metals ~\cite{Bard}, 
which was the starting point of  innumerable applications and theoretical developments in several fields.
 In the context of relativistic field theories important examples are vector 
dominance in strong-electromagnetic interactions~\cite{Gell} and dominance of chiral mesons in QCD. In  these cases the dominant composite 
bosons have fermion number zero, but in QCD at high density so called colour superconductivity~\cite{Color} is expected to occur,
and in this case  the dominant composite bosons  have fermion number 2  like  Cooper pairs in superconducting metals. Obviously
there is an immense literature about  these subjects, but since we address the problem of composite boson
dominance in its generality without actual applications to any of the above examples, we quote only historical works. We note however, that in spite 
of a huge literature about bosonization
all these systems are,  so far,  mostly treated phenomenologically. 

Nonrelativistic methods can hardly be extended  to relativistic field theories with few exceptions, one being 
ref.~\cite{Frol} which has some features in common with the relativistic papers quoted below.
We therefore only mention Bogoliubov's  work on superconductivity for his historical 
importance as the first organic approach to bosonization~\cite{Bogo}. We do not discuss
 fermion systems in 1+1 dimensions either. For them  a wealth of exact results exists which are however
 dimension specific and can be found in textbooks~\cite{Abda}.  

For relativistic  theories in higher dimensions the problem of bosonization has essentially two features: 
 how to introduce
the  fields associated to the  dominant composite bosons and how to handle the resulting effective action. There are basically two approaches. In the first one higher dimensional terms (quartic in the fermion fields) if absent, are introduced by hands in the action and the bosonic fields are generated by the Hubbard-Stratonovich transformation~\cite{Mira}. 
Sometimes, when the added terms are irrelevant in the renormalization-group terminology, they are introduced to stabilize the evaluation, for instance in numerical simulations, of the fermion determinant~\cite{Kogu}, but, more often, the investigations are analytical studies and are substantially restricted to a nonrenormalizable framework~\cite{Mira}. In one noticeable exception~\cite{Gies} the Hubbard-Stratonovich transformation is scale-dependent  ("re-bosonization") leading to a functional renormalization-group analysis.
In the second approach the boson fields are Kalb-Ramond fields introduced by a change of fermion  variables in the action~\cite{Burg}, a technique also adopted in the quoted work~\cite{Frol} about non relativistic many-body systems.
 No general procedure however has been developed
to handle the  fermionic  determinant appearing in the effective action, 
in particular for gauge theories. Moreover the above approaches are essentially restricted to composite bosons with fermion number zero.

Recently a  method of bosonization was developed in the framework of many-body theories by
 which we can  treat both charged and neutral composites ~\cite{Palu}.
The starting point in this method is the partition function in operator
form, namely the trace of the transfer matrix in the Fock space of the 
fermions. The physical assumption
of boson dominance is then implemented by restricting the trace to fermion
composites. This requires an
approximation of  a projection operator in the subspace of the composites, 
the approximation being the better,
the higher the number of fermion states in the composites. The approximate 
projection operator is constructed
in terms of coherent states of composites, and evaluation of the trace, which is done exactly, 
generates  a bosonic action in terms
of the holomorphic variables appearing in the coherent states.

Our approach shares two features with a variational method: The reduction of the starting full space, here the fermionic Fock space,  to a subspace, here that of the composites, and a variational procedure to determine the structure functions.  The  utility of variational methods and bosonization has been widely appreciated in the theory of many-body systems. But their potentiality has also been considered in the framework of relativistic field theories, in particular gauge theories, for for example by R.~Feynman~\cite{Feyn} who, however, was sceptical about its practical applicability, and recently in connection with QCD at high baryon density~\cite{Wilc}.

The particular bosonization  method we investigate has been completely developed for nonrelativistic many-body systems and checked 
on the BCS model
of superconductivity and the pairing model of finite systems like atomic nuclei and ultra-small
superconducting grains~\cite{Palu}. The properties of these systems   
are exactly reproduced. But in the nonrelativistic 
derivation of the effective
action, higher powers of the temporal spacing, which would not contribute in the 
continuum limit, were neglected.
This obviously cannot be done in relativistic theories because of ultraviolet 
divergences and in the present paper we  perform a different derivation.

We adopt  a lattice regularization for two reasons. The first one is that it 
allows an unambiguous definition of composites. The second one is related to gauge 
theories. We expect that in these theories the action of effective bosons will
involve vacuum expectation values of  invariant functions of gauge fields
which cannot be evaluated in the present framework. A lattice formulation should
allow us to extract such expectation values from numerical simulations.

There is a  price to be paid for such advantages related to  the well 
known difficulty with chiral invariance on a lattice, which can only in part be overcome by 
using Kogut-Susskind fermions. However our method can, at least in principle,
be used with any regularization for which a transfer matrix has been explicitly
constructed.
 
A major difference with respect to nonrelativistic theories is related to
Euclidean invariance. The formalism of the transfer matrix 
does not treat time and space in a symmetric way, and therefore Euclidean
invariance of the bosonic action must  be checked a posteriori. All other
symmetries are instead respected in our procedure.

We tested the validity of our method on  a model with 4-fermion interaction~\cite{Gros} in 3+1 dimensions: we exactly reproduce all the known results in the boson sector, namely condensation of a composite boson which breaks the discrete chiral
invariance of the model and its mass.
In addition we determine its structure function, whose spatial part is identical with that
of the Cooper pairs of the BCS model of superconductivity. Our approach in this case has some points 
in common with a variational calculation performed for the model in 1+1 dimensions~\cite{Mitc}.

Our solution of the model is more complicated than the standard one, but it is not
done looking for a greater simplicity, but as a check with special attention to 
Euclidean invariance. In any case  it gives explicitly the structure function of the
composite, which to our knowledge was not known.

In this work we confine ourselves to fermion systems at zero fermion density. An extension to non-vanishing fermion density will be presented elsewhere~\cite{Palu2}. Its application to the 4-fermion interaction model reproduces the properties of the fermion sector.

The paper is organized in the following way. In Section II we report, for the 
convenience of the reader, the general formalism. In Section III we derive 
an effective action for an arbitrary relativistic theory, and
discuss the saddle point approximation.
In Section IV we present an alternative derivation of the effective action with the relative saddle point approximation which turns out to be identical to the previous one. 
In Section V we apply our method to the model with 4-fermion interaction and in Section VI we summarize our 
results. Most technical details are relegated in Appendices.

\section{General formalism}

For  reader's convenience  we report the general formalism of 
bosonization based on coherent states of fermion composites. 
By using discrete time,  the operator form of the partition function is
\begin{equation} 
\mathcal{Z}_F =\mathrm{Tr}^F\, \prod_t \mathcal{T}_{t,t+1}\, .  \label{zf}
\end{equation}
The trace is over the Fock space of the fermions and   $\mathcal{T}_{t,t+1}$
is the Euclidean transfer matrix which maps the Fock space at time $t$ into that at time $t+1$.
 $t$ runs in a range which depends on the temperature.

Under the assumption of boson dominance we can restrict the trace to fermion 
bosonic composites. The restricted partition function can be written
\begin{equation}
\mathcal{Z}_C =\mathrm{Tr}^F\, \prod_t \mathcal{P} \,
\mathcal{T}_{t,t+1} \, ,     
\label{pf:trace}
\end{equation}
where $\mathcal{P}$ is a projection operator in the subspace of the 
composites. Because the formalism treats asymmetrically time and space, it
is convenient to use the following notation: we shall use boldface letters,
as $\mathbf{x}$, to denote spatial coordinates, and italic
letters to denote space-time coordinates: $x=(t,\mathbf{x})$.

To construct the projector $\mathcal{P}$ we first introduce the composite
creation operators
\begin{eqnarray}
\hat{\Phi}_{\mathbf{x},K}^\dagger=\hat{u}^\dagger\Phi_{\mathbf{x}K}^\dagger
\hat{v}^\dagger=
\sum_{ij}\hat{u}^\dagger_i(\Phi_{\mathbf{x}K}^\dagger)_{ij}\hat{v}^\dagger_j\, ,
\end{eqnarray}
where $\mathbf{x}$ represents the spatial coordinate of
the composite and $K$ its  quantum numbers, among which can be 
radial excitations, orbital angular momentum, spin, flavour, etc.
The $\Phi_{\mathbf{x}K}$ are the composite structure functions 
({\it wave functions}) and  
$\hat{u}^\dagger_i$ and $\hat{v}^\dagger_i$ are, respectively, creation 
operators of fermions and antifermions in state $i$, obeying canonical
anti-commutation relations,
\begin{equation}
\{\hat{u}^\dagger_i,\hat{u}_j\}=\{\hat{v}^\dagger_i,\hat{v}_j\}=\delta_{ij}
\, ,\; 
\{\hat{u}_i,\hat{u}_j\}= \{\hat{v}_i,\hat{v}_j\}=\{\hat{u}_i,\hat{v}_j\}=
\{\hat{u}^\dagger_i,\hat{v}_j\}=0 \, .
\end{equation}

Since the fermion creation operators are nilpotent, the composite creation operators $\hat{\Phi}^\dagger$
can be classified according  to their index of nilpotency, which is the highest integer exponent $\Omega$ such that 
\begin{equation}
\left (\hat{\Phi}^\dagger\right)^\Omega \neq 0\, .
\end{equation}

{\em The composite structure functions $\Phi_{\mathbf{x}K}$ are to be determined variationally in order to maximize the saturation of the partition function $\mathcal{Z}_C$.}
 
It is useful to introduce the operator doublet 
\begin{equation}
\hat{\psi} = \left(\begin{array}{c} \hat{u} \\ \hat{v}^\dagger \end{array}\right)
\end{equation}
and the orthogonal  projectors
\begin{eqnarray}
P_0^{(-)} \hat{\psi} & = & \hat{u} \\
P_0^{(+)}\hat{\psi} & = & \hat{v}^\dagger
\end{eqnarray}
in such a way that
\begin{equation}
\hat{\Phi}^\dagger = \hat{\psi}^\dagger P_0^{(-)} \Phi^\dagger P_0^{(+)} \psi\, .
\end{equation}

By analogy with canonical bosonic systems, we build coherent states of 
composites 
\begin{equation}
|\xi\rangle =
\exp\left(\sum_{\mathbf{x},K}\xi^{\vphantom{dagger}}_{\mathbf{x}K}
\hat{\Phi}^\dagger_{\mathbf{x}K}\right)\,|0\rangle\, ,
\end{equation}
where the $\xi_{\mathbf{x}K}$'s are  holomorphic variables and  $|0\rangle$ is the fermion vacuum.
We call these states coherent 
 because they have the form of  coherent states of elementary bosons,  sharing with these
states the property of a fixed phase relation among the components with different number of bosons. 
But the basic property of coherent states cannot be fulfilled. Indeed
\begin{equation}
\hat{\Phi}_{\mathbf{x}K} | \xi \rangle \neq \, \xi_{\mathbf{x}K} | \xi \rangle
\end{equation}
because composite creation-destruction operators don't satisfy canonical commutation rules.
Again in analogy with canonical bosonic systems, we define the operator
\begin{equation}
\mathcal{P}=\int  \left[ { d\xi d\xi^* \over 2 \pi i} \right]\,
\frac{1}{\langle\xi|\xi\rangle} \, |\xi\rangle \langle\xi| \, ,
\label{projector}
\end{equation}
where 
\begin{equation}
\left[ { d\xi d\xi^* \over 2 \pi i} \right]  =\prod_{\mathbf{x},K} 
\left[ { d\xi^{\vphantom{*}}_{\mathbf{x}K} d\xi^*_{\mathbf{x}K} \over 2\pi i } \right]
\end{equation}
which is neither a projector nor the identity in the subspace of the composites, but, as shown in Appendix~\ref{B}, it approximately becomes a projector onto the space of composites when the latter ones have a large index of nilpotency.
Indeed since
the composites do not obey canonical commutation relations, their properties
 can be very different from those of canonical bosonic
coherent states. However, if the index of nilpotency of the composites
is large enough, the composite system resembles a canonical bosonic system,
and  all the properties of canonical boson coherent states 
will approximately hold for the composite coherent states.

The scalar product of coherent states is
\begin{equation}
\langle\xi|\xi'\rangle =
{\det}_{+} \, (\mathcal{I}\:+\:B\,B'^\dagger) \label{detb}
\end{equation}
where
\begin{equation}
B^\dagger = \xi\cdot\Phi^\dagger = \sum_{\mathbf{x},K}\xi_{\mathbf{x}K}^{\vphantom{\dagger}}\, \Phi_{\mathbf{x}K}^\dagger \label{Bm}
\end{equation}
and for any matrix $\Lambda$ we define
\begin{equation}
{\det}_{\pm} \Lambda := \det (P_0^{(\pm)} \Lambda)\,. \label{det}
\end{equation}
The projection operator $P_0^{(\pm)} $ appears when matrices act on half the fermion field. 
{\it Notice that  the  entries  of all the matrices do not include  time}.
 $\mathcal{I} $ is the identity in the space of these matrices. {\it By a little abuse of notation we will 
 write "1" instead of  $\mathcal{I} $ when there should be no ambiguity in the interpretation. Similarly
 we will replace by "1" the identity in various subspaces like color, taste and so on}.

\section{ First form of the effective boson action}


According to~\cite{Lusc},  the fermion part of the transfer matrix, which is a function of  all elementary bosonic (also gauge) fields, that we represent by $\sigma$,   is factorized as
\begin{equation}
\mathcal{T}_{t,t+1} = \mathcal{T}\left[\sigma_t, \sigma_{t+1}\right] =\hat{T}^\dagger\left[\sigma_t \right] \hat{T}\left[\sigma_{t+1}\right] =  \hat{T}^\dagger_t \, \hat{T}_{t+1}^{\vphantom{\dagger}}\,  .
\end{equation}
We use the subscript $t$ to denote the dependence of any matrix on the particular configuration of bosonic fields $\sigma_t$, and
\begin{equation}
\hat{T} = \exp [ -\hat{u}^{\dagger} M  \, \hat{u} - \hat{v}^{\dagger}  
M^T \hat{v} ] 
\exp[\hat{v} N  \, \hat{u}]
\end{equation}
aside from a possible extra factor which is a function only of the bosonic fields and can be therefore included in the bosonic contribution to the partition function.
The  form  of the matrices $M$ and $N$ (the superscript $T$ means transposed) depends on the nature 
of the interactions and the regularization adopted for the fermions. 
What follows does not depend on their explicit expressions which are reported 
in the Appendix~\ref{A} for  Kogut-Susskind fermions in the flavour basis (we do not know any  suitable expression 
in the spin-diagonal basis). 

At finite time-strip of length $L_0$, with periodic boundary conditions for the bosonic fields, the partition function~(\ref{zf}) is
\begin{equation}
\mathcal{Z}_F = \mathrm{Tr}^F\, \hat{T}^\dagger_0\, \hat{T}_{1}^{\vphantom{\dagger}}\,\hat{T}^\dagger_1 \cdots \hat{T}^\dagger_{L_0-1}\,\hat{T}_{0}^{\vphantom{\dagger}}\,
\end{equation}
while its restriction to the composites~(\ref{pf:trace}) is
\begin{eqnarray}
\mathcal{Z}_C &=& \mathrm{Tr}^F\, {\cal P}\,\hat{T}^\dagger_0\, \hat{T}_{1}^{\vphantom{\dagger}}\,{\cal P}\,\hat{T}^\dagger_1 \cdots {\cal P}\,\hat{T}^\dagger_{L_0-1}\,\hat{T}_{0}^{\vphantom{\dagger}}\nonumber\\
&=& \int \prod_{t=0}^{L_0-1} \left[ \frac{d\xi_t d\xi^*_t}{2 \pi i} \right] \frac{1}{\langle \xi_t | \xi_t \rangle} 
\langle \xi_t | \hat{T}^\dagger_t \, \hat{T}_{t+1}^{\vphantom{\dagger}} | \xi_{t+1} \rangle \label{zcn} 
\end{eqnarray}
where we have introduced a copy of the Fock space of the composites at each time slice.
Explicitly
\begin{equation}
|\xi_t \rangle =
\exp\left(\sum_{\mathbf{x},K}\xi_K\left(t,\mathbf{x}\right){\vphantom{dagger}}
\hat{\Phi}^\dagger_{\mathbf{x}K}\left[\sigma_t\right]\right)\,|0\rangle\, .
\end{equation}
{\em We remark that the structure functions $\Phi$ do not depend explicitly on time, but as they are defined in presence of an external bosonic field configuration, time will enter as a label of the bosonic fields.}

We introduce also copies of the matrix $B$, defined in~(\ref{Bm}) at each time slice
\begin{equation}
B^\dagger_t  = \left(\xi\cdot\Phi^\dagger\right)_t = \sum_{\mathbf{x},K}\xi_{K}^{\vphantom{\dagger}}(t,\mathbf{x})\, \Phi_{\mathbf{x}K}^\dagger\left[\sigma_t\right] \, .
\end{equation}
Now setting  $M=0$, a restriction which will be eliminated in the next Section, we
evaluate  the matrix elements 
\begin{equation}
\langle \xi_t | \hat{T}^\dagger_t \, \hat{T}_{t+1}^{\vphantom{\dagger}} | \xi_{t+1} \rangle = \langle 0 | e^{v B_t u} e^{u^\dagger N_t^\dagger v^\dagger} e^{v N_{t+1} u} e^{u^\dagger B_{t+1}^\dagger v^\dagger} |0\rangle\, .
\end{equation}
With the help of the formulae collected in 
Appendix~\ref{nc} we find 
\begin{eqnarray}
\lefteqn{\langle 0 | e^{v B_t u} e^{u^\dagger N_t^\dagger v^\dagger} e^{v N_{t+1} u} e^{u^\dagger B_{t+1}^\dagger v^\dagger} |0\rangle}\nonumber\\
&& = \int \left[\frac{d\alpha^* d\alpha d\beta^* d\beta }{\langle\alpha\beta|\alpha\beta\rangle} \right] \langle 0 | e^{v B_t u} e^{u^\dagger N_t^\dagger v^\dagger} |\alpha\beta\rangle\langle \alpha\beta |e^{v N_{t+1} u} e^{u^\dagger B_{t+1}^\dagger v^\dagger} |0\rangle\nonumber\\
&& ={ \det}_{+} (1+ B_{t+1}^*N_{t+1}^T)\, {\det}_{+} (1 + B_t N_t^\dagger) \,\nonumber\\
&&  \hphantom{=} \times \int [d\alpha^* d\alpha d\beta^* d\beta]e^{-\alpha^*\alpha -\beta^* \beta - \beta^*\frac{1}{1+B_{t+1}^* N_{t+1}^T} B_{t+1}^* \alpha^* - \alpha B_t^T \frac{1}{1 + N_t^* B_t^T} \beta} \nonumber\\
&& =  {\det}_{+} (1+ B_{t+1}^*N_{t+1}^T)\, {\det}_{+} (1 + B_t N_t^\dagger) \,\nonumber\\
&&  \hphantom{=} \times { \det}_{-}\left( 1+ B_{t+1}^\dagger \frac{1}{1+N_{t+1} B_{t+1}^\dagger}\frac{1}{1+B_t N_t^\dagger} B_t \right)\nonumber\\
&&  = {\det}_{+} \, D_{t,t+1} 
\end{eqnarray}
where 
\begin{equation}
D_{t,t+1} =  \left(1+B_t N_t^\dagger\right) \left(1 + N_{t+1}B_{t+1}^\dagger\right) +  B_t B_{t+1}^\dagger\label{D}\, .
\end{equation}

Therefore by substitution in~(\ref{zcn}), using equation~(\ref{detb})
\begin{equation}
\mathcal{Z}_C =\int \prod_t \left[ { d\xi_t d\xi_t^* \over 2 \pi i} \right]\,
\exp\left[-S_\mathrm{eff}(\xi^*,\xi)\right]\, 
\end{equation}
where
\begin{eqnarray} 
\prod_t \left[ { d\xi_t d\xi_t^* \over 2 \pi i} \right] &=& \prod_{x,K} 
\left[ { d\xi_K(x) d\xi^*_{K}(x) \over 2\pi i } \right] \\
S_\mathrm{eff} &=& \sum_t \mathrm{tr}_{+}\,\left[  \ln \left(1 + B_t B_t^\dagger\right) -  \ln D_{t,t+1} \right]
 \label{S1}\,.
\end{eqnarray}
In analogy to (\ref{det}) we used the definition  valid for  any matrix $\Lambda$ 
\begin{equation}
\mbox{tr}_{\pm} \,\Lambda := \mbox{tr} (P_0^{(\pm)}\Lambda)\,.
\end{equation}
and we have replaced $(t,\mathbf{x})$ by $x$   to shorten the notation.

It is remarkable that the effective action for the composites $S_\mathrm{eff} $ has been evaluated exactly, so that 
{\it the 
only approximations in the partition function are the physical assumption of boson dominance  and the
form of the projector over the subspace of the composites.}

Remember  that
 the entries of these matrices  do not not include  time, which appears only as a label. For instance the matrix 
 $D$ satisfies the equation
\begin{equation}
D_{t,t+1} =  D_{t+1,t}^\dagger \label{simm}\, ,
\end{equation}
where  the  operation of Hermitean conjugation does not affect time.

\subsection{The saddle point equations}

When the index of nilpotency  is large the partition function is dominated by the minimum of the action. Its variation with respect 
to the matrix elements of $B$ and $B^{\dagger}$ gives
\begin{eqnarray}
B_t^\dagger \frac{1}{1 + B_t B_t^\dagger} & = & \left[ N_t^\dagger \left( 1 + N_{t+1} B_{t+1}^\dagger \right) + B_{t+1}^\dagger \right] \frac{1}{D_{t,t+1}} \label{first}\\
\frac{1}{1 + B_{t+1} B_{t+1}^\dagger} B_{t+1} & = & \frac{1}{D_{t,t+1}} \left[ \left(1 + B_t N_t^\dagger \right) N_{t+1}+ B_t\right]  \,. \label{second}
\end{eqnarray}
As a consequence of~(\ref{simm}), these equations are not independent from each other, because 
 the second  can be obtained from the first one  by exchanging  $t$ with $t+1$ after Hermitean conjugation. So it is sufficient to study only one of them.
 
 
In the sequel  we consider the case in which the effective action has a minimum with respect to the bosonic fields, coupled with the fermions, at a constant value. Of course, this assumption cannot hold for gauge fields. We are then able to determine the solutions
which provide the composite structure functions at the semiclassical level.
 
If the bosonic fields appearing in $N_t$ are constant at the saddle point,  there are constant nontrivial solutions  
$B_t = \overline{B}$ for the second order equation
\begin{equation}
\overline{B} N^\dagger \overline{B} - \overline{B} N^\dagger N  - N =0 \,  .
\end{equation}
Setting
 \begin{equation}
\overline{B} = N A
\end{equation} 
 for $A\neq 0$ we find
\begin{equation}
N\,\left(A\, N^\dagger N  A- A\, N^\dagger N  - 1\right) =0 \, .
\end{equation}
If we separate $A$ into its Hermitean and anti-Hermitean parts, we see from this equation that $A$ commutes with $N^\dagger N$.
The solution
\begin{equation}
A_{\pm}=\frac{  H \pm\sqrt{ 1 + H^2}}{2 H} \, , \label{sol}
\end{equation}
where 
\begin{equation}
H^2={ 1 \over 4} N^{\dagger} N \, , \label{H2}
\end{equation}
shows that $A$ is Hermitean.

 Remark that 
\begin{eqnarray}
A_{\pm} &=& - \left[A_{\mp} N^\dagger N\right]^{-1} \label{ca}
\\
A_+ & +&  A_-  = 1 \,.\label{unit}
\end{eqnarray}
The effective action at the saddle points
\begin{eqnarray}
\overline{S}_{\pm} & = & - \sum_t \mbox{tr}_{+}\, \ln \left( 1 + N A_{\pm}N^\dagger \right) \\
& = &
- \sum_t \mbox{tr}_{-}\, \ln \left( A_{\pm}^2 N^\dagger N \right) \\
& = &
- \sum_t \mbox{tr}_{-}\,\ln \left(- \frac{A_{\pm}}{A_{\mp}}\right) \\
&=& - \sum_t \mbox{tr}_{-} \ln \left( \sqrt{1+H^2} +H\right)^{\pm2}
\end{eqnarray}
takes opposite values, the  minimum being for the upper sign. The  ground state is a condensate of composites whose structure 
function in polar representation is
\begin{equation}
\overline{B} ={ 1\over 2} N  H^{-1}  f(H),
\end{equation}
 the polar radius being
\begin{equation}
f(H) =  \sqrt{ 1 + H^2} + H .
\end{equation}
In such composites  the occupancy of high momentum fermion states is larger than that of low momentum states. 
Such a structure of condensed bosons is quite different from the structure of (even virtually) bound pairs.
But it can be set in a more natural form by the unitary transformation 
\begin{equation}
(\hat{v}')^{\dagger} = \hat{u}, \qquad (\hat{u}')^{\dagger}=\hat{v} \,.  \label{unitary}
\end{equation}
It defines a new vacuum
\begin{equation}
\hat{u}' | 0 ' \rangle = \hat{v}' | 0 ' \rangle =0 
\end{equation}
related to the original vacuum according to
\begin{equation}
| 0'  \rangle =  \prod_i \hat{u}_i ^{\dagger} \hat{v}_i ^{\dagger}|0 \rangle\, . 
\end{equation}
{\it The new vacuum is the trivial solution of the saddle point equations. It is  the  completely filled state and it is then physically
equivalent to the original vacuum.} 

In the next Section we will derive a new form of the effective action using the new creation-annihilation operators.

\section{Second form of  the effective boson action}

We perform the evaluation of the effective action with the new operators. 
We first remark that the transformation (\ref{unitary}) interchanges the role of the projectors $P_0^{(\pm)}$. 
Since the eigenstates of these projectors correspond  to fermions propagating forward, respectively backward in time, this 
transformation is related to time-reversal. Under its action
\begin{equation}
 \hat{\psi}^\dagger P_0^{(+)} N P_0^{(-)} \hat{\psi} =   \hat{\psi}'^\dagger P_0^{(-)} N P_0^{(+)} \hat{\psi}'\, .
 \end{equation}
 In the general case this transformation induces the replacements
 \begin{eqnarray}
N & \leftrightarrow & N^\dagger \nonumber\\
\Phi &  \leftrightarrow & \Phi^\dagger \nonumber \\
M &  \leftrightarrow & - M  \label{rimp}
\end{eqnarray}
and a change in the purely bosonic contribution to the action.
At $M=0$ it interchanges in form $\hat{T}$ with $\hat{T}^\dagger$ whenever $N$ is Hermitean.
Now~(\ref{zcn}) becomes
\begin{eqnarray}
\mathcal{Z}_C &=& \mathrm{Tr}^F\, {\cal P}\,\hat{T}^{\vphantom{\dagger}}_0\, \hat{T}_{1}^\dagger\,{\cal P}\,\hat{T}^{\vphantom{\dagger}}_1 \cdots {\cal P}\,\hat{T}^{\vphantom{\dagger}}_{L_0-1}\,\hat{T}_{0}^\dagger\nonumber\\
&=& \int \prod_{t=0}^{L_0-1} \left[ \frac{d\xi_t d\xi^*_t}{2 \pi i} \right] \frac{1}{\langle \xi_t | \xi_t \rangle} 
\langle \xi_t | \hat{T}^{\vphantom{\dagger}}_t \, \hat{T}_{t+1}^\dagger\ | \xi_{t+1} \rangle \label{zcn2} 
\end{eqnarray}
where we have used the same notation as before for the coherent states but now
\begin{equation}
| \xi_{t} \rangle = \exp  \left( \xi_{t} \,\hat{v}' \Phi^{\dagger} \hat{u}' \right) 
\prod_j \left[(\hat{u}'_j) ^{\dagger} (\hat{v}'_j )^{\dagger} \right]|0' \rangle\, .
\end{equation}
To construct the effective action  we have  to evaluate the matrix elements
\begin{eqnarray}
\langle \xi_t | \hat{T}^{\vphantom{\dagger}}_t \, \hat{T}_{t+1}^\dagger | \xi_{t+1} \rangle& = & \langle 0'| \left[ \prod_i (\hat{u}'_i) ^{\dagger} (\hat{v}'_i )^{\dagger}\right]^{\dagger}
 \exp \left( \xi_t^*  (\hat{u}')^{\dagger} \Phi (\hat{v}')^{\dagger}  \right) 
 \hat{T}_t^{\vphantom{\dagger}} \, \hat{T}^{\dagger}_{t+1}
\nonumber\\ 
 & & \times  \exp  \left( \xi_{t+1} \hat{v}' \Phi^{\dagger} \hat{u}' \right) 
\prod_j \left[(\hat{u}'_j) ^{\dagger} (\hat{v}'_j )^{\dagger} \right]|0' \rangle \,.
\end{eqnarray}
Now we notice that for an elementary boson the basis of coherent states
\begin{equation}
 \exp ( b \, \hat{b}) (\hat{b}^\dagger)^{\Omega}|0 \rangle \, ,\,\,\, \exp ( b \, \hat{b}^{\dagger})|0 \rangle \,, 
\end{equation}
are equivalent to each other in the subspace of states with $0,1,\ldots,\Omega$ bosons. By analogy 
instead of the states $\exp  ( \xi \hat{v}' \Phi^{\dagger} \hat{u}' )$
$\times \prod_i (\hat{u}'_i)^{\dagger} (\hat{v}'_i )^{\dagger} |0' \rangle $ we can use the  basis 
$ \exp  \left(  \xi (\hat{u}') ^{\dagger}\Phi  \, (\hat{v}')^{\dagger}  \right)|0' \rangle $.
They are not exactly equivalent because the composites do not satisfy canonical commutation relations, but they 
should be  equivalent within our approximations, and we will have a crosscheck of this assumption. 
In the following we drop the prime on all creation-destruction operators. 

First we evaluate the action of $\hat{T}^{\dagger}$ on coherent states
\begin{equation}
\hat{T}^\dagger | \xi \rangle = \exp[\hat{u}^\dagger N^\dagger  \, 
\hat{v}^\dagger] \exp [ -\hat{u}^{\dagger} M^\dagger  \, \hat{u} - \hat{v}^{\dagger}  
M^* \hat{v} ]  \exp [\hat{u}^{\dagger} B^\dagger \hat{v}^{\dagger}] |0 \rangle \, .
\end{equation}
By repeated application of the identity, valid for arbitrary matrices $A, B$
\begin{equation}
\exp [ \hat{u}^{\dagger} B  \, \hat{u} ] 
 \exp [ \hat{u}^{\dagger} \,A \, \hat{v}^{\dagger} ] |0 \rangle = 
  \exp [ \hat{u}^{\dagger} \, e ^B \, A \, \hat{v}^{\dagger} ] |0 \rangle
\end{equation}
we get
\begin{equation}
\hat{T}^\dagger |\xi \rangle 
= \exp  \left\{ \hat{u}^{\dagger} \,
 \left[ N^\dagger +  e^ {-M^\dagger} \,B^\dagger  \, e^{-M^\dagger}\right] \,\hat{v}^{\dagger} \right\} |0 \rangle \,.
\end{equation}
Then the matrix elements of the transfer matrix are 
\begin{eqnarray}
\langle\xi_t |  \, \hat{T}^{\vphantom{\dagger}}_t \, \hat{T}_{t+1}^\dagger \, |\xi_{t+1}\rangle &=&
{\det}_{+}\left\{ 1\:+\:[N_{t} + \exp (-M_{t}) B_t \exp (-M_{t})] 
\right. 
\nonumber\\
& & \left. \times  [N_{t+1}^\dagger
+ \exp (-M^{\dagger}_{t+1}) B^\dagger_{t+1} \exp (- M^{\dagger}_{t+1})]
\right\} 
\end{eqnarray}
and  the new effective action reads
\begin{eqnarray}
S'_\mathrm{eff}  &=&  \sum_t
\mbox{tr}_{+}\left\{ \ln  \left[ 1 + B_t B^\dagger_t \right]
-  \ln   \left[  1 +   \left( N_t+ \mathrm{e}^{M_t}\, B_t\,
 \mathrm{e}^{M_t}\right) \, \right. \right.
 \nonumber\\
& & \left. \left. \times 
 \left(  N_{t+1}^\dagger
+ \mathrm{e}^{M^\dagger_{t+1}} \,B^\dagger_{t+1}\, 
\mathrm{e}^{ M^\dagger_{t+1}} \right)  \right]  \right\} \nonumber
\label{seff}
\end{eqnarray}
in which we have changed the sign of the matrix $M$ according to~(\ref{rimp}).
Notice that while the partition function must not change by application of the unitary transformation (\ref{unitary}),
the effective action need not remain the same.
For our application to the model with 4-fermion  interaction we are interested in the case of $M_t=0$. In the following we shall restrict ourselves to this case, then
\begin{equation}
S'_\mathrm{eff} = \sum_t \mbox{tr}_{+}\, \left[\ln \left(1 + B_t B_t^\dagger\right) -
 \ln {D}'_{t,t+1}\right] \label{S2}
\end{equation}
with
\begin{equation}
D'_{t,t+1} = 1 + \big(N_t+B_t\big)\left(N_{t+1}^\dagger + B^\dagger_{t+1}\right)
\end{equation}
Let us investigate what is the relation between this effective action and the one we got in the previous section. 
Its expression (\ref{S1}) can be rewritten as
\begin{eqnarray*}
S_\mathrm{eff} &=& \sum_t \mbox{tr}_{+}\, \left\{ \ln \left[B_t \left(B_t^{-1} (B_t^\dagger)^{-1} + 1\right) 
B_t^\dagger \right]  \right.\\
&& \left. -  \ln \left\{B_t \left[ \left(B_t^{-1} + N_t^\dagger\right) \left((B_{t+1}^\dagger)^{-1} +
 N_{t+1}\right) + 1 \right]B_{t+1}^\dagger \right\} \right\}\\
&=& \sum_t \mbox{tr}_{-}\, \left\{ \ln \left(1+ B_t^{-1} (B_t^\dagger)^{-1} \right) \right.\\
&&\left. -  \ln  \left[1+  \left(B_t^{-1} + N_t^\dagger\right) \left((B_{t+1}^\dagger)^{-1} + N_{t+1}\right) 
 \right] \right\}
\end{eqnarray*}
which coincides with $S'_{\mbox{eff}}$ under the change
\begin{equation}
B_t \mapsto (B_t^\dagger)^{-1} = B_t \left[ B_t^\dagger B_t\right]^{-1} \label{cb}
\end{equation}
for every $t$.
%
Under this change the saddle-point equation for time independent solutions  $B_t = \overline{B} = N A$ becomes
\begin{equation}
\left(A^{\dagger} \,N^\dagger N  A^\dagger + N^\dagger N  A^\dagger - 1\right)\, N^\dagger =0 \, . \label{sad}
\end{equation}
This is the same as the equation we had in the previous section under the change $A\to - A$, therefore we have the 
time independent solutions 
\begin{equation}
\overline{B}_{\pm}
 = - N A_{\pm}= \mp {N \over 2H} (\sqrt{ 1+H^2} \pm H) \,. 
\end{equation}
We see that in the saddle point approximation $\Phi$ is Hermitian when $N$ is Hermitian.
At the saddle point
\begin{eqnarray}
\overline{S}'_{\pm} 
& = &
- \sum_t \mbox{tr}_{-}\, \ln \left( 1 + A_{\mp}\,N^\dagger N \right) \\
& = &
- \sum_t \mbox{tr}_{-}\, \ln \left(-  \frac{A_{\mp}}{A_{\pm}} \right)\\ 
&=& - \sum_t \mbox{tr}_{-}\ln \left( \sqrt{1 +H^2}  +H  \right)^{\mp2}
\end{eqnarray}
takes its minimum for the lower sign, and
\begin{equation}
\overline{S}_+ = \overline{S}'_-.
\end{equation}
{\it At the minimum the values of the  two actions coincide, and therefore the corresponding partition functions are equal
at leading order}.
Let us notice that the transformation (\ref{cb}), because of the identity~(\ref{ca}), interchanges $A_+$ with $-A_-$.

The trivial and nontrivial solutions are strikingly similar to the corresponding solutions of the BCS model of superconductivity.
The comparison is best done using a polar representation for the structure functions which separates the unitary factor which 
depends on the intrinsic degrees of freedom, from the polar factor which is a function of the spatial coordinates. Now the unitary
factors  are necessarily different, because different are the intrinsic degrees of freedom, but the polar factors and therefore the spatial wave functions 
turn out to be identical. Indeed in the case of superconductivity
 the trivial solution is given by $\xi_k =\infty$ for $k < k_F$, $\xi_k = 0$, for 
$k_k  > k_F$, $k$ being the momentum and $k_F$ the Fermi momentum of the electrons while  the structure function of the condensed 
Cooper pairs is given by \cite{Palu}
\begin{equation}
\sigma_2 \left( \sqrt{1 + E^2} - E \right) \,.
\end{equation}
 The Pauli matrix $ \sigma_2$ which couples the spins to zero is replaced in a relativistic theory by the spin-taste structure matrix
 $ N (2 \,H)^{-1} $
and   the electron kinetic energy $E$ measured with respect  to the chemical potential divided by the energy gap is replaced by the relativistic energy  $H$. 
In the present case the chemical potential disappears, because the composites are neutral, and  to investigate the effect of a 
chemical potential one should introduce states
with non vanishing fermion number. 
What makes the comparison with the BCS model somewhat involved is the absence of a Fermi energy in relativistic theories
with zero chemical potential. 
Indeed the fully occupied state in the old operators
of the relativistic theory corresponds to the state where all single particle states are occupied below the Fermi surface in the BCS model.

\section{Application to a model with 4-fermion interaction}

We apply our formalism to the field theory with quartic interaction in 3+1 dimensions regularized
on a lattice with Kogut-Susskind fermions. 
 For each of the four Kogut-Susskind tastes we take $N_f$
degenerated flavours. Hence, the continuum limit will describe a theory
with $4N_f$ flavours. 
In the flavour basis the action reads 
\begin{equation}
\mathcal{S}=  \sum_x{}^\prime \sum_y{}^\prime \bar\psi(x) \,\left[ m \, 1\!\!1  \otimes 1\!\!1  \, + Q \right]_{x,y} \psi(y) + 
{ 1 \over 2}\,  { g^2 \over 4 N_f} \sum_x{}^\prime (\bar\psi(x)\psi(x))^2  \, 
\end{equation}
where $m$ is the mass parameter, $g^2$ the coupling constant,  $\psi$ the fermion fields and 
 $Q$  the  hopping matrix:
\begin{equation}
Q= \sum_\mu\gamma_\mu\otimes 1\!\!1
\left[ P_\mu^{(-)}\nabla_\mu^{(+)} + P_\mu^{(+)}\nabla_\mu^{(-)}\right]\, .
\end{equation}
The matrices to the left (right) of the symbol $\otimes$ act on Dirac (taste) indices. We denote by $\gamma$ and $t$ the matrices acting on 
these indices, respectively. The operators
\begin{equation}
P_\mu^{(\pm)}=\frac{1}{2}
\left[1\!\!1 \otimes 1\!\!1 \pm\gamma_\mu\gamma_5\otimes t_5t_\mu\right]\, .
\label{spin_projectors}
\end{equation}
are orthogonal projectors. The fermion fields are defined on blocks (see Appendix~\ref{A} for details). The right
and left derivatives $\nabla^{(\pm)}$ are given by 
\begin{equation}
\nabla^{(\pm)}_{\mu} = \pm { 1 \over 2} \left( T^{(\pm)}_{\mu} - 1  \right) \,.
\end{equation}
The factor $ 1/2$ is due to the fact that the $T_{\mu}$ translate by one block
\begin{equation}
\left( T^{(\pm)}_{\mu}  \right)_{x_1, x_2}= \delta_{x_2, x_1 \pm 2 \hat{\mu}}  \, .
\end{equation} 
The model has a discrete chiral symmetry at $m=0$:
\begin{equation}
\psi\,\rightarrow\, -\gamma_5\otimes 1\!\!1 \,\psi\, , \hspace{1truecm}
\bar\psi\,\rightarrow\, \bar\psi\,\gamma_5\otimes 1\!\!1 \, .
\end{equation}
To have an action bilinear in the fermion fields we introduce auxiliary 
scalar field $\sigma(x)$, whose integration generates the 
four fermion coupling:
\begin{equation}
\mathcal{S}'= \sum_x{}^\prime \sum_y{}^\prime \bar\psi(x) \,  
(m+\sigma+Q)_{xy}\psi(y) + \frac{4 N_f}{2g^2}\sum_x{}^\prime \,\sigma^2(x)  \, .
\end{equation}
The partition function now reads
\begin{equation}
\mathcal{Z}=\int[d\sigma][ d \overline{\psi} d \psi ] \, \exp\left[- \mathcal{S}' \right] \, .
\end{equation}

\subsection{The effective boson action} 

In the model  with 4-fermion interaction with Kogut-Susskind fermions in the flavour basis~\cite{Palu1} the matrix $M=0$, while the matrix $N$ is Hermitian and is given by 
\begin{equation}
N(\sigma) =
-2  \left\{ \left( m + \sigma \right)\, \gamma_0  \otimes 1\!\!1  +  \sum_{j=1}^3  \gamma_0  \gamma_j   \otimes 1\!\!1  
 \left[  P^{(-)}_j   \nabla_j^{(+)}  + P^{(+)}_j \nabla_j^{(-)}
\right] \right\}  \, .
\end{equation}
We will restrict ourselves to flavour singlet
composites, what means that $\Phi_{\bf{x} K}$ acts trivially on flavour indices, 
and thus, obviously, only flavour singlet composites can be written as 
linear combination of the $\Phi_{\bf{x} K}$. From now on, we will ignore flavour
indices. 
According to the results of the above Sections, the effective action at the saddle point is
\begin{eqnarray}
S_\mathrm{eff}(\bar\xi^*,\bar\xi,\overline{\sigma})  &=&
 - \sum_t \hbox{tr}_{-}  \ln \left[ \sqrt{1 + H^2} + H  \right]^2 \\ 
&=& - { L_0 \over 2} \, \mbox{tr}
\ln \left[ \sqrt{1 +H^2} +H \right] \label{azione}
\end{eqnarray}
because the factor $1/2$ coming from the projector $P_0^{(-)}$, present in $\hbox{tr}_{-}$,  is compensated by the exponent 2 in the $\ln$. 

We now show that this is equal to the standard result. The latter is obtained by a direct integration over the fermion fields in the partition function
\begin{equation}
\mathcal{Z}= \int [d\sigma] \exp \left[ - S_F  -\frac{4\,N_f}{2g^2}\sum_{x}{}^\prime\, \sigma^2(x)\right]
\end{equation}
where
\begin{eqnarray}
S_F &=&  - \mbox{Tr} \ln  (m + \sigma+Q) \\
& = & -\frac{L_0}{2}  \, \mbox{tr} \,\int_{-\pi/2}^{\pi/2} \frac{d\omega}{2 \pi} \ln \left[H^2 + \frac{1}{2} \left(1 - \cos 2 \omega\right)\right] \, .
\end{eqnarray}
Here $\mbox{Tr}$ (notice the capital case T) is the trace on all the entries of $Q$, namely time is included, and
since $Q$ acts on the whole fermion field there is no projector $P_0^{(-)}$. The explicit integration over $\omega$ reproduces, apart from a constant term, the expression in~(\ref{azione}). 
Therefore 
\begin{equation}
{\partial \over \partial \overline{\sigma}}S_\mathrm{eff}(\bar\xi^*,\bar\xi,\overline{\sigma})  
=  {\partial \over \partial \overline{\sigma}} S_F
= - { L_0 \over 2} \, \mbox{tr}
\, { \overline{\sigma}\over H\sqrt{ 1 +  H^2} } \,.
\end{equation}

\subsection{Mass of the effective boson}

Let us consider the Gaussian fluctuations around the minimum of the action.
To this end, let us write
\begin{eqnarray}
\xi_K(x) &=& \overline{\xi}_K\:+\:\varphi_K(x)\, , \\
\xi_K^*(x) &=& \overline{\xi}_K^*\:+\:\varphi_K^*(x)\, , \\
\sigma(x) &=& \overline{\sigma}\:+\:\eta(x)\,. 
\end{eqnarray}
We can assume the $ \overline{\xi}_K$ real, because the $\xi_K$ are defined up to 
a global phase which can be eliminated by a redefinition of the $\Phi_{\mathbf{x}K}$.

Let us expand the effective action 
$S_\mathrm{eff}(\overline{\xi}+\varphi^*,\overline{\xi}+\varphi,\overline{\sigma}+\eta)$ 
to second order in powers of $\varphi$, $\varphi^*$, and $\eta$. 
The linear term  vanishes due to the gap equation and the 
terms involving $\varphi  \, \varphi$ and $\varphi^*\varphi^*$ vanish on 
the saddle point.
Ignoring the constant term 
$S_\mathrm{eff}(\overline{\xi}^*, \overline{\xi},\overline{\sigma})$, 
the quadratic part of the effective action  reads
\begin{eqnarray}
S_\mathrm{eff} &=& 
\sum_{x,x'}\,\eta(x)\,D^{(\eta\eta)}(x,x')\,\eta(x')
\nonumber \\
&&+ \sum_{xK,x'K'}\,
\varphi_K^*(x)\,D^{(\varphi^*\varphi)}_{KK'}(x,x')\,\varphi_{K'}(x')
\nonumber \\
&&+ \sum_{x,x'K}\,\eta(x)\, D^{(\eta\varphi)}_K(x,x')\,\varphi_K(x')
\nonumber \\
&&+\sum_{x,x'K}\,\varphi_K^*(x)\, D^{(\varphi^*\eta)}_K(x,x')\,\eta(x')
\end{eqnarray}
where the $D$'s are given by traces involving the  matrices  
$\Phi_{\mathbf{x}K}$, $\bar\Phi_{\mathbf{x}K}$, and 
$N$ (see appendix~\ref{C} for details). Recall that the entries of these
matrices are the fermion spatial coordinates and  Dirac-taste indices, and 
the label $K$ of the composite bosons includes Lorentz, taste and spatial 
(radial and angular momentum) quantum numbers. 
We expand the structure matrices $\Phi_{\mathbf{x}K}$ in a basis of matrices 
$\Xi(\mathbf{x},\mathbf{n})$ which act only on the spatial coordinates of 
the fermions and  
$\Gamma_{\gamma}$ which act only on the Dirac-taste indices
\begin{equation}
\Phi_{\mathbf{x}K}  = \sum_{\mathbf{x}_1 \mathbf{n}\gamma} 
c_{\mathbf{x}K,\mathbf{x_1}\mathbf{n}\gamma} \, 
\Xi(\mathbf{x_1},\mathbf{n})\,\Gamma_\gamma\, .
\end{equation}
We accordingly expand the boson fields
\begin{equation}
\varphi_{ \mathbf{n},\gamma}(t, \mathbf{x}) = \sum_{\mathbf{x}_1 K} c_{\mathbf{x}_1K, \mathbf{x} \mathbf{n}\gamma} \, 
\varphi_K(t,\mathbf{x}_1) \, .
\end{equation}
We consider two options for the matrices $ \Xi(\mathbf{x},\mathbf{n})$ 
\begin{equation}
\left[\,\Xi(\mathbf{x},\mathbf{n})\,\right]_{\mathbf{x}_1, \mathbf{x}_2}\;=\;
\delta_{\mathbf{x}_1, \mathbf{x} + { 1 \over 2} \mathbf{n}}\,
\delta_{\mathbf{x}_2, \mathbf{x} - { 1 \over 2} \mathbf{n}}\, ,
\end{equation}
\begin{equation}
\left[\,\Xi(\mathbf{x},\mathbf{n})\,\right]_{\mathbf{x}_1, \mathbf{x}_2}\;=\;\delta_{\mathbf{x}_1,\mathbf{x}} 
\,\delta_{\mathbf{x}_2,  \mathbf{n}}.
\label{def:Xi}
\end{equation}
In the first case $\mathbf{x}$ has the physical meaning of the centre of mass coordinate of the composites and $\mathbf{n}$ of the 
relative distance of the fermions. Then the  matrix $\Xi $ must be separated 
into two contributions:
one for  $\mathbf{x}= 2\mathbf{r}_x, \mathbf{n}=4 \mathbf{ r}_n$, the other one for
$\mathbf{x}=2( \mathbf{r}_x+1), \mathbf{n} = 4\mathbf{ r}_n +2$, writing $\mathbf{x} = 2\mathbf{ r}$, where the $\mathbf{r}$ are
vectors with integer components. This accounts for the fact that 
the coordinate of the center of mass can fall in the full lattice, while the 
positions of the fermions and their relative distance are coordinates of blocks. 
In the second case $\mathbf{x}$ and $\mathbf{x}$ are  the coordinates of  the fermions.

For the matrices $\Gamma$ we require
\begin{eqnarray}
\mathrm{Tr}\, (\Gamma_\gamma^\dagger\,\Gamma_{\gamma'}) &=& 
\delta_{\gamma\gamma'} \, .
\end{eqnarray}
Note that  the $\Gamma_\gamma$'s verify the identity
\begin{equation}
P_0^{(-)}\,\Gamma_\gamma\,P_0^{(+)} = \Gamma_\gamma\, .
\end{equation}
If we are interested in the dynamics of the $\varphi$-fields we must determine 
the structure matrices and therefore the coefficients $c$. But in this paper 
we have the limited goal of reproducing the spectrum of the bosonic composite $\eta$.
For this purpose we integrate out the $\varphi$-fields and we do not need to 
determine the coefficients $c$.

Integration of the $\varphi$-fields generates  an effective action for
$\eta$ of the form
\begin{equation}
S_\mathrm{eff}[\eta] = \sum_{x,x'}\,\eta(x)\,\left[D^{(\eta\eta)} + K^{(\eta\varphi)} \,\left(K^{(\varphi^{*}\varphi)}\right)^{-1}\,\left(K^{(\eta\varphi)}\right)^{\dagger}\right](x,x')\,\eta(x') \,.
\end{equation} 
The explicit expression of the kernels appearing in this expression can be found in Appendix~\ref{C}, but we remark that they don't depend from the structure functions of the $\varphi$-fields. 
Close to the continuum limit ($\overline{\sigma}\rightarrow 0$),
its  behaviour is given by 
\begin{eqnarray} 
S_\mathrm{eff}[\eta] &\simeq & \int_{-\pi/2}^{\pi/2} \frac{d\omega}{2\pi}
\int_{-\pi/2}^{\pi/2} \frac{d^3 \mathbf{p}}{(2\pi)^3}
\,\frac{1}{2}\,
\eta(-\omega,-\mathbf{p}) \,\eta(\omega,\mathbf{p}) \nonumber \\
&& \times \left[\,Z_\omega(\overline{\sigma}^2)\, \omega^2 
+ Z_p(\overline{\sigma}^2)\, \sum_i \mathbf{p}_i^2 +
W(\overline{\sigma}^2)\,
\overline{\sigma}^2\,\right]\, .
\end{eqnarray}
As $\overline{\sigma}\rightarrow 0$, the integrals defining
$Z_\omega(\overline{\sigma}^2)$, 
$Z_p(\overline{\sigma}^2)$, and
$W(\overline{\sigma}^2)$ as discussed in Appendix~\ref{C}, are dominated by the 
logarithmic infrared divergence, and we get
\begin{equation}
Z_\omega(\overline{\sigma}^2) \simeq
Z_p(\overline{\sigma}^2) \simeq 
\frac{W(\overline{\sigma}^2)}{4} \simeq -
\frac{1}{(2 \pi)^2}\,\ln \overline{\sigma}^2 \:+\: O(1) \, .
\end{equation}
The logarithmic divergences are the same which appear in the exact 
solution of the model at large $N_f$, leading to logarithmic corrections to scaling
and thus to the triviality of the continuum limit.
In conclusion, the mass of the scalar is
\begin{equation}
m_\eta \;=\; 2 \overline{\sigma} \, ,
\end{equation}
which is the exact relation with the chiral
condensate obtained from the gap equation.
Notice also that Euclidean invariance is recovered in the continuum limit:
\begin{equation}
\frac{Z_\omega(\overline{\sigma}^2)}{Z_p(\overline{\sigma}^2)}\;\rightarrow\; 1
\, .
\end{equation}

Summarizing, in this Section we have shown that the hypothesis of boson
dominance and the associated formalism, applied to the 4-fermion interaction model
reproduces the exact solution in the boson sector, namely, the correct gap 
equation, mass gap, and  logarithmic corrections to scaling.

{\it We know that the spectrum of the model contains also one
fermion of mass $m_\psi=\overline{\sigma}$. This property is reproduced by an extension of the present formalism}~\cite{Palu2}.

\section{Summary and outlook}

In this paper we applied  a general method of bosonization to relativistic 
field theories whose low energy excitations are dominated by bosonic modes. 
This  is always the case in the presence of spontaneous breaking of continuous 
symmetries. One of our goals  is indeed the study of low energy hadron dynamics
in QCD.

Under the condition of boson dominance the fermion partition function can be 
restricted to boson composites. To realize this
restriction we use the formalism of the transfer matrix, which  is  close  to 
the Hamiltonian formalism 
of nonrelativistic theories, and therefore very natural dealing with real or 
virtual bound states.
The projection onto the subspace of composites in the partition function is 
realized introducing coherent  states of composites.
The projection operator is not exact, but its  approximation is the more 
accurate the higher  the index of nilpotency of the composites (the number 
of fermionic states which define their structure functions). Indeed it leads 
to exact results in the model with quartic interaction in the $ N_f  \rightarrow \infty $ 
limit, in which case the index of nilpotency also tends to infinity.
Evaluation of traces in the fermion Fock space can be performed without any 
further approximation  for theories which are quadratic in the fermion fields or 
can be put in such form, a condition satisfied by all renormalizable models 
in 3+1 dimensions. 
It generates a functional form of the partition function 
with an effective action of the composites.
In this derivation 
time and space are not treated in the same way, and Euclidean invariance must be checked
a posteriori, but  all other symmetries are respected including 
of course gauge invariance. 

We checked  our theory  on the 4-fermion interaction model in 3+1
dimensions. Not only did we reproduce all the known results in the boson sector, 
but we have also been able to determine the structure functions of the condensed composite.
It turns out that the polar factor  of the structure
function is identical to that of  the Cooper pairs of the BCS model of superconductivity. 

The study of the 4-fermion interaction model in our formalism is not complete, 
because we did not investigate if, in addition to the sigma,  there are other bosonic modes. 
Indeed what we actually did was to prove that projecting onto the composites subspace and then integrating over
the fermionic fields gives the same results as a direct integration (without projection) of these fields as far as spontaneous breaking
of chiral invariance and  mass of the composite are concerned. We must also observe that while the gap equation does
not depend on the number of space-time dimensions, our evaluation of the mass of the sigma is restricted to 3+1
dimensions because we used completeness relations valid in this space.

The effective action we derived can be used in the study of QCD by numerical simulations and, we hope, by a perturbative expansion along the lines of~\cite{Palu} and by keeping into account the subtleties related to the $1/N_f$ expansion~\cite{Caracciolo}. In this case the effective bosons carry obviously the quantum numbers of the chiral mesons and chiral symmetry is broken by the condensation of the sigma-meson.
The introduction since the beginning of the expectation value of the sigma-field as 
a variational parameter, should help in a numerical simulation, as advocated in Ref.~\cite{Kogu}. But we must remember that the form factors of the 
composites depend in general on the fields of the
elementary bosons coupled to the fermions, as we have explicitly seen in the  model with 4-fermion interaction. For an actual numerical simulation
in QCD therefore, we need a trial expression of the meson form factors, which should 
depend on the gauge fields, as also required by gauge-invariance of the effective action. Apart from the interest
which it has per se, a perturbative approach might prove useful also to provide trial expressions of the form factors.

We conclude by observing that the present formulation can be immediately used also for studies of systems at finite temperature, while for finite fermion density we must first 
include  states with nonvanishing fermion number.

\section*{Acknowledgments}

This work has been partially supported by an INFN-CICyT collaboration and
by Ministerio de Educaci\'on y Ciencia 
(Spain), projects FPA2003-02948 and BFM2003-08532-C03-01/FISI.
V. Laliena is a Ram\'on y Cajal fellow.

F. Palumbo has been partially supported by EEC under the contract ``Forces Universe" MRTN-CT-2004-005104.

\appendix

\section{Some notations and conventions\label{nc}}
We have used the following identities valid for Berezin integrals 
\begin{eqnarray}
\int [d\alpha^* d\alpha]\, e ^{-\alpha^* \alpha} & = & 1\\
\int [d\alpha^* d\alpha]\, e ^{-\alpha^* A \alpha} & = & \det A\\
\int [d\alpha^* d\alpha]\, e ^{-\alpha^* A \alpha + J^*\alpha + \alpha^* J } & = & \det A \,
e^{J^* A^{-1} J}\, .
\end{eqnarray}
If $|\alpha\rangle$ is a fermionic coherent state
\begin{equation}
|\alpha\rangle = e^{-\alpha \hat{u}^\dagger} |0\rangle
\end{equation}
then
\begin{equation}
\langle\alpha|\alpha\rangle = e^{\alpha^* \alpha} \langle0|0\rangle = e^{\alpha^* \alpha}
\end{equation}
and
\begin{equation}
\int [d\alpha^* d\alpha]\, \frac{1}{\langle\alpha|\alpha\rangle} |\alpha\rangle\langle\alpha| = I\, .
\end{equation}
Remark that
\begin{equation}
\hat{u} \,|\alpha\rangle =  \alpha \, |\alpha\rangle 
\end{equation}
which implies the relations
\begin{eqnarray}
\langle\alpha\beta| e^{\hat{v}N\hat{u}}|\gamma\delta\rangle & = & e^{\delta N \gamma} \langle\alpha\beta|\gamma\delta\rangle =  e^{\delta N \gamma+\alpha^*\gamma +
\beta^*\delta}\\
\langle\gamma\delta|e^{\hat{u}^\dagger B^\dagger \hat{v}^\dagger}|0\rangle & = & \langle0| e^{\hat{v}B\hat{u}}|\gamma\delta\rangle^* = e^{\gamma^*B^\dagger \delta^*}\, .
\end{eqnarray}
With the help of these formulae we will compute
\begin{eqnarray}
\lefteqn{\langle \alpha \beta |e^{\hat{v} N \hat{u}} e^{\hat{u}^\dagger B^\dagger \hat{v}^\dagger} |0\rangle}\\
&& = \int \left[\frac{d\gamma^* d\gamma  d\delta^* d\delta}{\langle\gamma\delta|\gamma\delta\rangle}\right] \langle \alpha \beta |e^{\hat{v} N \hat{u}}|\gamma\delta\rangle\langle\gamma\delta| e^{\hat{u}^\dagger B^\dagger \hat{v}^\dagger} |0\rangle \\
&& = \int [d\gamma^* d\gamma d\delta^* d\delta] e^{-\gamma^*\gamma - \delta^*\delta +\delta N \gamma +\alpha^*\gamma + \beta^*\delta+\gamma^*B^\dagger\delta^*}\\
&& =  \int [d\delta^* d\delta] \, e^{ - \delta^*(1+B^* N^T)\delta  + \beta^*\delta - \delta^*B^*\alpha^*}\\
&& =  {\det}_{+} (1+B^* N^T)\, e^{-\beta^*\frac{1}{1+B^* N^T} B^* \alpha^*}
\end{eqnarray}

\section{ The operator $\mathcal{P}$ \label{B}}

The restriction of the partition function to the subspace of composite bosons can be written
\begin{equation}
Z_C = \mbox{Tr}^{\mbox{F}}  \prod_t  \mathcal{P}\,\mathcal{T}_{t,t+1} 
\end{equation}
where $ \mathcal{P}  $ is the projection operator on this subspace.
For the sake of simplicity we consider the case of a unique composite. In such a case 
\begin{equation}
 \mathcal{P}  = \sum_{n =0}^{\Omega}  { 1 \over  \nu_n} 
   |\left( \hat{\Phi}^\dagger \right)^n|0 \rangle \langle 0| \hat{\Phi}^n |
\end{equation}
where
\begin{equation}
 \nu_n  =  \langle 0| \hat{\Phi}^n | \left( \hat{\Phi}^\dagger \right)^n|0 \rangle \,.
\end{equation}
We must show that
\begin{eqnarray}
\langle 0 | \hat{\Phi}^m \, \mathcal{P} ( \hat{\Phi}^\dagger )^n |0\rangle 
\simeq  \langle 0 | \hat{\Phi}^m |  ( \hat{\Phi}^\dagger)^n |0\rangle  = \delta_{m,n} \nu_m \,.\label{Papp1}
\end{eqnarray}
These equations are generated by the following ones
\begin{equation}
\langle \xi' | \mathcal{P} | \xi'' \rangle \simeq \langle \xi' | \xi'' \rangle \label{eqnostra}
\end{equation}
whose right and left hand sides are
\begin{eqnarray}
\langle \xi' | \mathcal{P} | \xi'' \rangle & =& \int { d \xi^* d\xi \over 2 \pi i} 
\exp (-\mathcal{E(\xi^*,\xi, \xi'^*, \xi'')})
\nonumber\\
\langle \xi' | \xi'' \rangle &=& \exp \mbox{tr}_{+}\ln (1 + \xi'^* \xi'' \Phi \Phi^\dagger)
\end{eqnarray}
where
\begin{equation}
\mathcal{E}(\xi^*,\xi, \xi'^*, \xi'') =  \mbox{tr}_{+} \left[ \ln (1 + \xi^* \xi  \Phi \Phi^{\dagger}) 
-  \ln (1 + \xi'^* \xi  \Phi \Phi^{\dagger}) -  \ln (1 + \xi^* \xi''  \Phi \Phi^{\dagger})  \right] \,.
\end{equation}
We evaluate the integral by the saddle point method.  The saddle point equations are
\begin{eqnarray}
(\xi - \xi'')\, \mbox{tr}_{+}\, { \Phi \Phi^{\dagger} \over (1 + \xi^* \xi \Phi \Phi^{\dagger} )
(1 + \xi^* \xi'' \Phi \Phi^{\dagger} )} &=&0
\nonumber\\
(\xi^* - \xi'^*)\, \mbox{tr}_{+}\, { \Phi \Phi^{\dagger} \over (1 + \xi^* \xi \Phi \Phi^{\dagger} )
(1 + \xi'^* \xi \Phi \Phi^{\dagger} )} &=&0  
\end{eqnarray}
with solutions
\begin{equation}
\xi = \xi'' \,\,\, \xi^* = \xi'^* \,.
\end{equation}
At the saddle point
\begin{equation}
\mathcal{E}(\xi^*,\xi, \xi'^*, \xi'') =  -  \mbox{tr}_{+} \ln (1 + \xi'^* \xi''  \Phi \Phi^{\dagger}) \,.
\end{equation}
Moreover
\begin{eqnarray}
\frac{\partial^2 {\cal E}}{\partial \xi \partial \xi^*} & = & 
\frac{\partial^2 {\cal E}}{\partial \xi \partial \xi} \\
\frac{\partial^2 {\cal E}}{\partial \xi^* \partial \xi} & = & \mbox{tr}_{+}\, {  \Phi \Phi^{\dagger}
\over (1 + \xi'^* \xi''  \Phi \Phi^{\dagger})^2} \,.
\end{eqnarray}
In conclusion
\begin{equation}
\langle \xi' | \mathcal{P} | \xi' \rangle \simeq \langle \xi' | \xi'' \rangle  \, \left[ \mbox{tr}_{+}
 {  \Phi \Phi^{\dagger}
\over (1 + \xi'^* \xi''  \Phi \Phi^{\dagger})^2} \right]^{-1}\,.
\end{equation}
In  presence of a number of composites  $ n << \Omega$ the desired result, 
together with the idempotency property of projector 
\begin{equation}
 \mathcal{P} \simeq    \mathcal{P}^2 
\end{equation}
follow if we assume
\begin{equation}
\mbox{tr} \left( \Phi^\dagger \Phi \right)^n \sim \Omega^{-n+1} \, .
\end{equation}

But in  states with  $n  \sim \Omega$, it is not possible to satisfy~(\ref{eqnostra})
even with an absolute freedom about the form of the structure functions (which are instead determined by
the dynamics). The best we can do is to satisfy~(\ref{eqnostra}) (apart from an irrelevant constant factor) for states with $n+k$ composites, for fixed $n\sim \Omega$ and
 $|k| << \Omega$. 

To clarify this point let us go back to the case of many composites and consider their commutation relations
\begin{equation}
\left[ \hat{\Phi}_{\alpha} ,   \hat{\Phi}^{\dagger}_{\beta}  \right] = 
 \mbox{tr} (\Phi_{\alpha} \Phi^{\dagger}_{\beta}) 
- \hat{u}^{\dagger}\Phi^{\dagger}_{\beta} \Phi_{\alpha} \hat{u }
 - \hat{v}^{\dagger}\Phi^*_{\beta} \Phi^T_{\alpha} \hat{v }. \label{comm}
\end{equation}
In  states with a number of composites  $ n << \Omega$, these relations are approximately canonical provided the structure functions are sufficiently smooth.
Indeed in such a case the last 2 terms are of order $n / \Omega $.

\section{The transfer matrix for Kogut-Susskind fermions\label{A}}

Kogut-Susskind fermions in the flavour basis are defined on hypercubes of twice the ordinary lattice spacing. We shall have a fermionic field $\psi_i^{\alpha\, a}(t,\mathbf{x})$ (and of course the corresponding $\overline{\psi}$), where $i=\{1,\ldots,N_f\}$ is the flavour index, $\alpha=\{1,\ldots,4\}$ is the spinorial index, $a=\{1,\ldots,4\}$ is the {\em taste} index, while
$(t,\mathbf{x})$ is a 4-vector of {\em even} integer components ranging in the intervals $[0,L_t-1]$ for the time component while $[0, L_s-1]$ for each spatial components.

In order to simplify the notations let restrict here to periodic functions defined on one dimensional interval $x\in [0,L-1]$, the extension to many dimensions and anti-periodic functions being obvious. 

The sum over even sites $x$ includes for convenience a factor 2
\begin{equation}
\sum_x{}^\prime := 2 \sum_x  \,.
\end{equation}
Momenta are quantized according to
\begin{equation}
p = \frac{2 \pi}{L}\,n \qquad n = 0, 1, \frac{L}{2} -1 
\end{equation}
The Fourier series  and its inverse for the function $f(x)$ defined on even sites are
\begin{eqnarray}
\tilde{\tilde{f}}(p) &:=&  \sum_{x=0}^{L-2} {}^\prime f(x)\, \mathrm{e}^{i x p}\\
f(x) &:=& { 1 \over L} \sum_{n=0}^{L/2 -1} \tilde{\tilde{f}}\left( { 2 \pi \over L}\,n\right)  \mathrm{e}^{-  i \frac{ 2 \pi}{L}  n x}\,.
\end{eqnarray}
Notice that we use a double tilde to distinguish this transform  from the standard one for a function defined on all the sites which will be denoted by a single tilde.
As $f$ is a function of an even argument we could set
\begin{equation}
g(s) := f (2 s)
\end{equation}
where $s$ is an integer in the interval $[0,L/2-1]$, then their Fourier transforms are simply related as 
\begin{eqnarray}
\tilde{\tilde{f}}(p) & = & 2 \sum_{s=0}^{L/2-1} f(2s)  \, \mathrm{e}^{i 2 s p}\\
& = & 2 \sum_{s=0}^{L/2-1} g(s)  \, \mathrm{e}^{i 2 s p} \\
& = & 2 \, \tilde{g}(2 p)\vphantom{\sum_{s=0}^{L/2-1}}
\end{eqnarray}

In the infinite volume limit we get the Fourier transform
\begin{eqnarray}
\tilde{\tilde{f}}(p) &:=&  \sum_{x=0}^{\infty} {}^\prime f(x)\, \mathrm{e}^{i x p}\\
f(x) &:=&\int_{-\pi/2}^{\pi/2} \frac{dp}{2\pi} \,\tilde{\tilde{f}}(p)\,  \mathrm{e}^{- i x p }\,.
\end{eqnarray}

As a consequence, for instance, the $\delta$-function must be written
\begin{equation}
\delta(x-y) = 2 \int_{-\pi/ 2}^{\pi / 2} { dp\over 2\pi}\,\mathrm{e}^{i p (x-y)} 
\end{equation}
and the Fourier transform of a  matrix of the form
\begin{equation}
\Lambda = F(- \bigtriangleup)
\end{equation}
for an arbitrary function $F$ is 
\begin{equation}
\tilde{\tilde{\Lambda}}(p)= 2  F \left( { 1\over 2} ( 1 - \cos 2p) \right)
\end{equation}
and its trace
\begin{equation}
\mbox{tr} \, \Lambda= \frac{L }{2}\,\int_{-\pi/ 2}^{\pi / 2} { dp\over 2\pi} F\left( { 1\over 2} ( 1 - \cos 2p) \right)
 = \sum_{x}{}^\prime\int_{-\pi/ 2}^{\pi / 2} { dp\over 2\pi} F \left( { 1\over 2} ( 1 - \cos 2p) \right)\,.
\end{equation}

We report here the expression of the transfer matrix in the flavour basis~\cite{Palu1}, because the expression
in the spin-diagonal basis is not known to us in a convenient form.
The matrix $N$, which is Hermitean, in presence  of a gauge field denoted by $U$, is given by
 \begin{eqnarray}
\lefteqn{N(\sigma,U) =} \\
&&  -2  \left\{  (m+\sigma)  (\gamma_0  \otimes \one)  +  \sum_{j=1}^3  (\gamma_0  \gamma_j   \otimes \one)
 \left[  P^{(-)}_j   \nabla_j^{(+)}  + P^{(+)}_j \nabla_j^{(-)}
\right]  \vphantom{\sum_{j=1}^3} \right\}\nonumber
\end{eqnarray}
where
\begin{equation}
\nabla^{(\pm)}_j = \pm { 1\over 2} \left( \hat{U}^{(\pm 1)}_j (t)\, T_j^{(\pm)} - 1 \right) \,
\end{equation}
are right-left covariant derivatives.
The $\hat{U}(t)$'s are operators whose matrix elements are the Wilson link variables $U_{\mu}(t, {\bf x}_1) $
\begin{equation}
\left( \hat{U}_{\mu}(t) \right)_{ {\bf x}_1, {\bf x}_2} \,=\, \delta_{ {\bf x}_1, {\bf x}_2}\, U_{\mu}(t, {\bf x}_1).
\end{equation}
The matrix $M$ depends on the  link $U_0$ variable. In the gauge
$ U_0  \sim   1   $, in which $U_0 = 1$ with the exception of one time slice, $M=0$. If we adopt this gauge,
we must impose the Gauss constraint on the states. But it is very simple to write 
the expression of $M$ without gauge fixing following Ref.~\cite{Smit}. This is actually necessary in a numerical simulation.

In the absence of gauge fields  and with constant $\sigma=\overline{\sigma}$, according to (\ref{H2})
\begin{equation}
H^2 =  \left(  m +\overline{\sigma}  \right) ^2 -  \Delta\,.
\end{equation}
with
\begin{equation}
\Delta = { 1 \over 4}  \sum_{i=1,3} \left( T_i^{(+)}+ T_i^{(-)} -2  \right)
\end{equation}
The eigenvalues of $H^2  $ are therefore the fermion energies
\begin{equation}
E_q^2 = { \left(m+\overline{\sigma}\right)^2}  + \tilde{q}^2 \, , \label{energy}
\end{equation}
where the  momentum $\tilde{q}^2 $ is 
\begin{equation}
\tilde{q}^2 = \sum_{i=1}^3 \tilde{q}^2_i
\end{equation}
with
\begin{equation}
\tilde{q}^2_i = { 1\over 2} ( 1  - \cos 2q_i)\, . \label{momentum}
\end{equation}

\section{ Quadratic fluctuations \label{C}}

Let us give some details about the expansion of the effective action
around the constant fields $(\overline{B},\overline{B}^\dagger ,\overline{\sigma})$ which minimize it. We shall set
\begin{eqnarray}
B_t  & = & \overline{B} + \delta B_t \\
\sigma(t,\mathbf{x}) & = & \overline{\sigma}+\eta(t,\mathbf{x})
\end{eqnarray}
where 
\begin{eqnarray}
\overline{B} & = & \overline{\xi^* \cdot \Phi} \\
\delta B_t  & =  & (\varphi^* \cdot \Phi)_t = \sum_{\mathbf{x} K} \varphi_K(t,\mathbf{x})^* \Phi_{\mathbf{x}K}
\end{eqnarray}
and shall concentrate on quadratic fluctuations.

In the parametrization of the form-factors $\Phi_{\mathbf{x}K}$ we choose the  form in~(\ref{def:Xi})  for the matrix $\Xi(\mathbf{x},\mathbf{x'})$.

It is also convenient to introduce the matrix $\hat\eta_t$ by
\begin{equation}
\hat\eta_t \;=\; 2\,\sum_\mathbf{x}
\eta(t,\mathbf{x})\,\Xi(\mathbf{x},\mathbf{x})\,(\gamma_0\otimes \one)\, . 
\end{equation}

\subsection{Expansion  of the action}

According to our definitions the expansion for the effective action reads:
\begin{eqnarray}
\lefteqn{S_\mathrm{eff}(\varphi^*,\varphi,\eta) =  
\overline{S}_\mathrm{eff}
+\sum_t\,\mbox{tr}_{+}\left\{\,
\,R\,(\overline{B}\delta B_t^\dagger +
\delta B_t \overline{B}^\dagger) \right.}\nonumber \\
&& -\vphantom{\frac{1}{2}}\,R_N(\overline{B}_N\delta B_{t+2}^\dagger +
\delta B_t \, 
\overline{B}_N^\dagger + \overline{B}_N\hat\eta_{t+2}^\dagger
+\hat\eta_t \overline{B}_N^\dagger) \nonumber \\
&&+ \vphantom{\frac{1}{2}}
\,R\,\delta B_t \,\delta B_t^\dagger 
\,-\,R_N\,(\delta B_t \,\delta B_{t+2}^\dagger  
+\hat\eta_t\hat\eta_{t+2}^\dagger+
\hat\eta_t\,\delta B_{t+2}^\dagger +
\delta B_t \,\hat\eta_{t+2}^\dagger)
\nonumber \\
&&- \frac{1}{2}\,
[\,R\,(\overline{B}\,\delta B^\dagger \,+\,
\delta B_t \,\overline{B}^\dagger)\,]^2 \nonumber \\
&&+ \left. \frac{1}{2}\,
[\,R_N\,(\overline{B}_N\,\delta B_{t+2}^\dagger +
\delta B_t \,\overline{B}_N^\dagger+
\overline{B}_N\hat\eta_{t+2}^\dagger+\hat\eta_t 
\overline{B}_N^\dagger)\,]^2\,\right\}\, , 
\end{eqnarray}
where 
\begin{eqnarray}
\overline{B}_N & =& \overline{N} + \overline{B}\\
R & = & \left(1 + \overline{B}\overline{B}^\dagger\right)^{-1} \\
R_N & = & \left(1 + \overline{B}_N\overline{B}_N^\dagger\right)^{-1} \, .
\end{eqnarray} 
Ignoring the constant term, the expansion can be cast in the form
\begin{eqnarray}
S_\mathrm{eff} &=& 
\sum_{t\mathbf{x},t'\mathbf{x'}}\,\frac{1}{2}\,\eta(t,\mathbf{x})\,\eta(t',\mathbf{x'})\,D^{(\eta\eta)}(t,\mathbf{x};t',\mathbf{x'})
\nonumber \\
&&+\sum_{t\mathbf{x}K,t'\mathbf{x'}K'}\,
\varphi_K^*(t,\mathbf{x})\,\varphi_{K'}(t',\mathbf{x'})\,D^{(\varphi^*\varphi)}_{KK'}(t,\mathbf{x};t',\mathbf{x'})
\nonumber \\
&&+ \sum_{t\mathbf{x},t'\mathbf{x'}K}\,\eta(t,\mathbf{x})\,\varphi_K(t',\mathbf{x'})\, D^{(\eta\varphi)}_K(t,\mathbf{x};t',\mathbf{x'})
\nonumber \\
&&+\sum_{t\mathbf{x},t'\mathbf{x'}K}\,\eta(t,\mathbf{x})\,\varphi_K^*(t',\mathbf{x'})\, 
D^{(\eta\varphi^*)}_K(t,\mathbf{x};t',\mathbf{x'}) \, ,
\end{eqnarray}
since the linear terms vanish due to all stationarity equations 
---we are expanding around a minimum--- and the terms involving
$\varphi_K \varphi_{K'}$ and $\varphi_{K}^* \varphi_{K'}^*$ vanish
due to the stationarity equations for the fields $B$ and $B^\dagger$.

The $D$'s entering the above equations are:
\begin{eqnarray}
D^{(\eta\eta)}(t,\mathbf{x};t',\mathbf{x'}) &=& 4\,\mbox{tr}_{+}\,R_N\,\left\{\,
-2\,\delta_{t',t+2}\,\Xi(\mathbf{x},\mathbf{x})(\gamma_0\otimes \one)
\Xi(\mathbf{x'},\mathbf{x'})(\gamma_0\otimes \one)  \right. 
\nonumber \\
&+&
\delta_{t',t+2}\,
\Xi(\mathbf{x},\mathbf{x})(\gamma_0\otimes \one)\, \overline{B}_N^\dagger
R_N\overline{B}_N\Xi(\mathbf{x'},\mathbf{x'})(\gamma_0\otimes \one) 
\nonumber \\
&+&
\delta_{t,t'+2}\,
\overline{B}_N\Xi(\mathbf{x},\mathbf{x})(\gamma_0\otimes \one) \,
R_N\Xi(x',x')(\gamma_0\otimes \one) \overline{B}_N^\dagger \nonumber \\
&+&\delta_{tt'}\,[\overline{B}_N
\Xi(\mathbf{x},\mathbf{x})(\gamma_0\otimes \one)\, R_N\overline{B}_N
\Xi(\mathbf{x'},\mathbf{x'})(\gamma_0\otimes \one) 
\nonumber \\
&+& 
\left. 
\Xi(\mathbf{x},\mathbf{x})(\gamma_0\otimes \one)\, \overline{B}_N^\dagger R_N
\Xi(\mathbf{x'},\mathbf{x'})(\gamma_0\otimes \one)\, \overline{B}_N^\dagger]\,\right\} \, 
\label{detaeta} \\ \nonumber \\
D^{(\varphi^*\varphi)}_{KK'}(t,\mathbf{x};t',\mathbf{x'}) &=&
\mbox{tr}_{+}\,\left[\delta_{tt'}\,R\,\Phi_{\mathbf{x}K}\left(\,1\,-\,
\overline{B}^\dagger R \overline{B}\,\right)\Phi^\dagger_{\mathbf{x'}K'}\right.
\nonumber \\
&&-\, \delta_{t',t+2}\,\left. R_N\,\Phi_{\mathbf{x}K}\,\left(\,1\,-\,
\overline{B}^\dagger R_N 
\overline{B}\,\right)\,\Phi^\dagger_{\mathbf{x'}K'}\right]\, 
\label{dphiphi} \\ \nonumber \\
D^{(\eta\varphi)}(t,\mathbf{x};t',K) &=& 2\,\mbox{tr}_{+}\,R_N\,\left\{\,
-\delta_{t',t+2}\,
\Xi(\mathbf{x},\mathbf{x})(\gamma_0\otimes \one)\, \Phi_{\mathbf{x}K}^\dagger \right. \nonumber \\
&+& \delta_{tt'}\,\overline{B}_N
\Phi_{\mathbf{x}K}^\dagger R_N\overline{B}_N\Xi(\mathbf{x},\mathbf{x})(\gamma_0\otimes \one)
\nonumber \\
&+& \left. \delta_{t',t+2}\,\overline{B}_N\Phi_{\mathbf{x}K}^\dagger
\Xi(\mathbf{x},\mathbf{x})(\gamma_0\otimes \one)\, \overline{B}_N^\dagger\,\right\}\, 
\label{detaphi} \\ \nonumber \\
D^{(\eta\varphi^*)}(t,\mathbf{x};t',K) &=& 2\,\mbox{tr}_{+}\,R_N\,\left\{\,
-\delta_{t,t'+2}\,\Phi_{\mathbf{x}K}
\Xi(\mathbf{x},\mathbf{x})(\gamma_0\otimes \one)  \right. \nonumber \\
&+& \delta_{tt'}\,\Phi_{\mathbf{x}K} \overline{B}_N^\dagger
R_N\Xi(\mathbf{x},\mathbf{x})(\gamma_0\otimes \one)\, \overline{B}_N^\dagger \nonumber \\
&+& \left. \delta_{t,t'+2}\,\Phi_{\mathbf{x}K} \overline{B}_N^\dagger
R_N \overline{B}_N\Xi(\mathbf{x},\mathbf{x})(\gamma_0\otimes \one) \,
\right\}\, .
\label{detaphistar}
\end{eqnarray}

\subsection{Evaluation of the traces}

We need to make explicit the meaning of the indices $K$'s : $K = (\mathbf{n}, \gamma)$,
so that
\begin{equation}
\varphi_K  \;\longrightarrow\; \varphi_{\mathbf{n}\gamma}\, .
\end{equation}
and therefore
\begin{equation}
\begin{array}{lcl}
D^{(\varphi^*\varphi)}_{KK'} & \longrightarrow & 
D^{(\varphi^*\varphi)}_{\mathbf{n}\gamma,\mathbf{n'}\gamma'} \\[3mm]
D^{(\eta\varphi)}_K & \longrightarrow & 
D^{(\eta\varphi)}_{\mathbf{n}\gamma} \\[3mm]
D^{(\eta\varphi^*)}_K & \longrightarrow &
D_{\mathbf{n}\gamma}^{(\eta\varphi^*)}\, .
\end{array}
\end{equation}

Now we can readily compute the traces entering equations 
\mbox{(\ref{detaeta})-(\ref{detaphistar})}. In order to present
the results, it is convenient to introduce additional definitions.

On the solution of the gap equation, the matrices $R$ and $R_N$
can be written as
\begin{eqnarray}
R &=& \frac{A+1}{2A+1} \\
R_N &=& \frac{A}{2A+1}
\end{eqnarray}
where
\begin{equation}
A=\frac{\sqrt{1+ H^2}-H}{2 H}\, .
\end{equation}
Obviously, $R+R_N=I$.
Notice that the following relations are verified:
\begin{eqnarray}
A^2 &=& (4H)^{-2}\:-\:A\, , \\
(1+A)^2 &=& (4H)^{-2}\:+\:1\:+\:A\, .
\end{eqnarray}

We also introduce the following notation:
\begin{eqnarray}
\Theta &=& R_N\,(I+A)\, , \\
S_i^{(s)} &=& \Theta\,(T_i^{(s)}-1)\, , \hspace{0.7truecm} s=+,-\, ,
\end{eqnarray}
and
\begin{eqnarray}
t^\gamma &=& \mathrm{tr}'_{-}\,[\,(\gamma_0\otimes \one)\,\Gamma_\gamma\,\,]
\, , \\
t_{is}^\gamma &=& \mathrm{tr}'_{-}\,[\,(\gamma_i\gamma_0\otimes \one)
\,P_i^{(-s)}\,\Gamma_\gamma\,\,]\, , \\
t_{is,js'}^\gamma &=& \mathrm{tr}'_{-}\,[\,
(\gamma_j\otimes \one)\, P_j^{(-s')}\,(\gamma_0\gamma_i\otimes \one) \, P_i ^{(-s)}\,
\Gamma_\gamma\,\,]\, 
\end{eqnarray}
where $\mathrm{tr}'_{-}$ is the trace on the Dirac-taste space.
Introducing the Fourier transforms
\begin{eqnarray}
\eta(t,\mathbf{x}) &=&\int \frac{d\omega}{2\pi} \int \frac{d^3\mathbf{p}}{(2\pi)^3}\,
\mathrm{e}^{i \omega t}\,\mathrm{e}^{i   \mathbf{p}\cdot \mathbf{x}}\,\eta(\omega,\mathbf{p})\, , \\
\varphi_{\mathbf{n}\gamma}(t,\mathbf{x}) &=& \int \frac{d\omega}{2\pi} \int \frac{d^3\mathbf{p}}{(2\pi)^3} \int \frac{d^3\mathbf{q}}{(2\pi)^3} \,
\mathrm{e}^{i \omega t}\,\mathrm{e}^{i  \mathbf{p}\cdot \mathbf{x}}\,\mathrm{e}^{i  \mathbf{q}\cdot \mathbf{n}}\,
\varphi_{\mathbf{q}\gamma}(\omega,\mathbf{p})
\end{eqnarray}
we get the following effective action:
\begin{eqnarray}
\lefteqn{S_\mathrm{eff} =\int \frac{d\omega}{2\pi} \int \frac{d^3\mathbf{p}}{(2\pi)^3}\, \frac{1}{2}\,D^{(\eta\eta)}(\omega,\mathbf{p})\,\eta(-\omega,-\mathbf{p})\,\eta(\omega,\mathbf{p})}
\nonumber \\
&&+ \int \frac{d\omega}{2\pi} \int \frac{d^3\mathbf{p}}{(2\pi)^3} \int \frac{d^3\mathbf{q}}{(2\pi)^3}\,
D^{(\varphi^*\varphi)}_{\mathbf{q},\gamma \gamma'}(\omega,\mathbf{p})\,
\varphi_{\mathbf{q}\gamma}^*(\omega,\mathbf{p})\,\varphi_{\mathbf{q}\gamma'}(\omega,\mathbf{p})
\nonumber \\
&&+ \int \frac{d\omega}{2\pi} \int \frac{d^3\mathbf{p}}{(2\pi)^3} \int \frac{d^3\mathbf{q}}{(2\pi)^3}\,D_{\mathbf{q}\gamma}^{(\eta\varphi)}(\omega,\mathbf{p})\,
\varphi_{\mathbf{q}\gamma}(\omega,\mathbf{p})\,\eta(-\omega,-\mathbf{p}) \nonumber \\
&&+ \int \frac{d\omega}{2\pi} \int \frac{d^3\mathbf{p}}{(2\pi)^3} \int \frac{d^3\mathbf{q}}{(2\pi)^3} \, 
D_{\mathbf{q}\gamma}^{(\eta\varphi^*)}(-\omega,\mathbf{p})\,
\varphi_{\mathbf{q}\gamma}^*(\omega,\mathbf{p})\,\eta(\omega,\mathbf{p})\, , 
\end{eqnarray}
where
\begin{eqnarray}
D^{(\eta\eta)}(\omega,\mathbf{p}) &=&  \frac{1}{g^2} \:+\:
64\,\int \frac{d^3\mathbf{q}}{(2\pi)^3} \,\left\{\,-\,\vphantom{\sum_i}
(2-\cos2\omega)R_N(\mathbf{q})R_N(\mathbf{p}-\mathbf{q}) \right. \nonumber\\ 
&& \:-\: 
\frac{1}{2}\sum_i[S_i^{(+)}(\mathbf{q})S_i^{(-)}(\mathbf{p}-\mathbf{q})+S_i^{(-)}(\mathbf{q})S_i^{(+)}(\mathbf{p}-\mathbf{q})] \nonumber\\
&& \left.+  4\,\overline{\sigma}^2\,\Theta(\mathbf{q})\Theta(\mathbf{p}+\mathbf{q} )
\,\vphantom{\sum_i}\right\}\, ,\\
D^{(\varphi^*\varphi)}_{\mathbf{q},\gamma \gamma'}(\omega,\mathbf{p}) &=&
\delta_{\gamma\gamma'}\,\left[\,
R\left(\mathbf{p}\right)\,R\left(\mathbf{q}\right)\:-\:
e^{i 2 \omega}\,
R_N\left(\mathbf{p}\right)\,R_N\left(\mathbf{q}\right)\,\right]\, , \vphantom{\sum_i}\\
D_{\mathbf{q}\gamma}^{(\eta\varphi)}(\omega,\mathbf{p}) &=& 2\,\left\{\,\vphantom{\sum_\pm}
t^\gamma \,\left[\,
4\,\overline{\sigma}^2\,\Theta\left(\mathbf{p}\right)\,
\Theta\left(\mathbf{q}\right)
\:-\:\mathrm{e}^{i 2 \omega}\,R_N\left(\mathbf{p}\right)\,
R_N\left(\mathbf{q}\right)\,
\right]\right. \nonumber \\
&-&
2\,\overline{\sigma}\,
\sum_{i,s=\pm 1}\,s\,t_{is}^\gamma\,\left[\,
\Theta\left(\mathbf{q}\right)\,S_i^{(s)}\left(\mathbf{p}\right)
\:+\:\Theta\left(\mathbf{p}\right)\,
S_i^{(-s)}\left(\mathbf{q}\right)\,\right]
\nonumber \\
&+&\left. 
\sum_{i,j}\sum_{s,s'=\pm 1}\,s\,s'\,t^\gamma_{is,js'}\,
S_i^{(-s)}\left(\mathbf{q}\right)\,
S_j^{(s')}\left(\mathbf{p}\right)\,\right\}\, .
\end{eqnarray}

\subsection{Effective action for the $\eta$-field}

The integrations over $\varphi_{\mathbf{q}\gamma}$ and $\varphi_{\mathbf{q}\gamma}^*$ give the following effective action for $\eta$:
\begin{equation}
S_\eta =  \int \frac{d\omega}{2\pi} \int \frac{d^3\mathbf{p}}{(2\pi)^3}\,\frac{1}{2}\,\eta(-\omega,-\mathbf{p})\,\eta(\omega,\mathbf{p})\,\left[
D^{(\eta\eta)}(\omega,\mathbf{p})\:+\:\Delta ^{(\eta\eta)}(\omega,\mathbf{p})\,\right]\, ,
\end{equation}
where
\begin{equation}
\Delta ^{(\eta\eta)}(\omega,\mathbf{p}) \;=\;
 \int \frac{d^3\mathbf{q}}{(2\pi)^3}\,
\frac{ D_{\mathbf{q}\gamma}^{(\eta\varphi)\,*}(-\omega,\mathbf{p}-\mathbf{q})\,
D_{\mathbf{q}\gamma}^{(\eta\varphi)}(\omega,\mathbf{p}-\mathbf{q})}
{R\left(\mathbf{q}\right)\,R\left(\mathbf{p}-\mathbf{q}\right)\,-\,
\mathrm{e}^{i 2 \omega}\,R_N\left(\mathbf{q}\right)\,
R_N\left(\mathbf{p}-\mathbf{q}\right)}\, .
\end{equation}

The sum over $\gamma$ appearing in the above equations can be performed using the following identities
\begin{eqnarray}
\sum_\gamma t^{\gamma\,*} t^\gamma &=& \mathrm{tr}'_{-}\,
\left[\,\one\otimes\one\,\right] \;=\; 8\, , \\
\sum_\gamma t^{\gamma\,*} t_{is}^\gamma &=& \mathrm{tr}'_{-}\,
\left[\,(\gamma_i\otimes \one)\, P_i^{(s)}\,\right] \;=\; 0\, , \\
\sum_\gamma t^{\gamma\,*} t_{is,js'}^\gamma &=& \mathrm{tr}'_{-}\,
\left[\,(\gamma_i\otimes \one)\, P_i^{(s)}\,
(\gamma_j\otimes \one)\, P_j^{(s')}\,\right]\nonumber \\
&=& 4\,\delta_{ij}\,\delta_{s,-s'}\, , \\[4 mm]
\sum_\gamma t^{\gamma\,*}_{is} t^\gamma_{js'} &=& \mathrm{tr}'_{-}\,
\left[\,(\gamma_j\otimes \one)\,P_j^{(s')}\,
(\gamma_i\otimes \one)\, P_i^{(-s)}\,
\,\right]\nonumber \\
&=& 4\,\delta_{ij}\,\delta_{ss'} \\[4 mm]
\sum_\gamma t_{is}^{\gamma\,*} t_{js',ks''}^\gamma &=& \mathrm{tr}'_{-}\,
\left[\,(\gamma_j\otimes \one)\, P_j^{(s')}\,
(\gamma_k\otimes \one)\, P_k^{(s'')}\,(\gamma_i\otimes \one)\, P_i^{(-s)}
\,\right]\nonumber \\
&=&  0\, , \\[4 mm]
\sum_\gamma t_{i_1s_1,j_1s'_1}^{\gamma\,*}
t_{i_2s_2,j_2s'_2}^\gamma &=& \nonumber \\
\lefteqn{\hspace{-3cm}= \mathrm{tr}'_{-}\,
\left[\,(\gamma_{i_2}\otimes \one)\, P_{i_2}^{(s_2)}\,
(\gamma_{j_2}\otimes \one)\,P_{j_2}^{(s'_2)}\, 
P_{j_1}^{(s'_1)}\,(\gamma_{j_1}\otimes \one)\,P_{i_1}^{(s_1)}\,
(\gamma_{i_1}\otimes \one)\,\right]}\\
\lefteqn{\hspace{-3cm}= 2\,\left(\,
\delta_{i_1j_1}\delta_{i_2j_2}\delta_{s'_1,-s_1}\delta_{s'_2,-s_2}
\,+\,\delta_{i_1i_2}\delta_{j_1j_2}\delta_{s_1s_2}\delta_{s'_1s'_2}
\,-\,\delta_{i_1j_2}\delta_{i_2j_1}\delta_{s_1s'_2}\delta_{s'_1s_2}
\,\right)\, .\nonumber}
\end{eqnarray}
which follow from the completeness  relations satisfied by the matrices $\Gamma_\gamma$'s which form a basis in the relevant
subspace of matrices acting on Dirac and taste indices.

With the above equalities it is easy to get the explicit form of 
$\Delta ^{(\eta\eta)}(\omega,\mathbf{p})$:
\begin{eqnarray}
\lefteqn{\Delta ^{(\eta\eta)}(\omega,\mathbf{p}) =}\\
&& \,-\,64\, \int \frac{d^3\mathbf{q}}{(2\pi)^3}\,
\frac{1}{R\left(\mathbf{q}\right)\,R\left(\mathbf{p}-\mathbf{q}\right)
\:-\:\mathrm{e}^{i 2 \omega}\,R_N\left(\mathbf{q}\right)\,
R_N\left(\mathbf{p}-\mathbf{q}\right)}
\nonumber \\
&& \times\left\{\,\vphantom{\sum_i}\left[\,4\,\overline{\sigma}^2\,
\Theta\left(\mathbf{q}\right)\,\Theta\left(\mathbf{p}-\mathbf{q}\right)
\:-\:\mathrm{e}^{i 2 \omega}R_N\left(\mathbf{q}\right)\,
R_N\left(\mathbf{p}-\mathbf{q}\right)\,\right]^2 \right. 
\nonumber \\
&& \hphantom{\otimes} \vphantom{\sum_i}-\:\left[\,4\,\overline{\sigma}^2\,
\Theta\left(\mathbf{q}\right)\,\Theta\left(\mathbf{p}-\mathbf{q}\right)
\:-\:\mathrm{e}^{i 2 \omega}\,
R_N\left(\mathbf{q}\right)\,R_N\left(\mathbf{p}-\mathbf{q}\right)\,\right]\,
\nonumber \\
&& \hphantom{\otimes}\times\,\sum_i\,\left[\,
S_i^{(+)}\left(\mathbf{q}\right)\,S_i^{(+)}\left(\mathbf{p}-\mathbf{q}\right)+
S_i^{(-)}\left(\mathbf{q}\right)\,S_i^{(-)}\left(\mathbf{p}-\mathbf{q}\right)\,
\right] 
\nonumber \\
&& \hphantom{\otimes}+\: \sum_j\,
S_j^{(+)}\left(\mathbf{q}\right)\,S_j^{(-)}\left(\mathbf{q}\right) \sum_i\,
S_i^{(+)}\left(\mathbf{p}-\mathbf{q}\right)\,S_i^{(-)}\left(\mathbf{p}-\mathbf{q}\right)\,
\nonumber \\
&& \hphantom{\otimes}+\: \left. 4\,\overline{\sigma}^2\,\sum_i\,
\left|\,\Theta\left(\mathbf{p}-\mathbf{q}\right)\,S_i^{(+)}\left(\mathbf{q}\right)
\:+\:\Theta\left(\mathbf{q}\right)\,
S_i^{(-)}\left(\mathbf{p}-\mathbf{q}\right)\,\right|^2
\,\right\} \nonumber
\end{eqnarray}
Now we know $D^{(\eta\eta)}(\omega,\mathbf{p})$ and
$\Delta ^{(\eta\eta)}(\omega,\mathbf{p})$ in terms of integrals of known functions
of momenta and $\overline{\sigma}$. 

\subsection{Low energy expansion}

Let us consider the expansion of $D^{(\eta\eta)}(\omega,\mathbf{p})$ and
$\Delta ^{(\eta\eta)}(\omega,\mathbf{p})$ in powers of $\omega$ and $\mathbf{p}$ up to
second order:
\begin{equation}
D^{(\eta\eta)}(\omega,\mathbf{p}) \:+\:
\Delta ^{(\eta\eta)}(\omega,\mathbf{p})\ \;\simeq\;
Z_\omega(\overline{\sigma}^2)\,\omega^2\:+\:Z_p(\overline{\sigma}^2)\,
 \mathbf{p}^2 \:+\: W(\overline{\sigma}^2)\, \overline{\sigma}^2\, .
\end{equation}
The gap equation implies that the above expression at $\omega=0$ and $\mathbf{p}=0$ 
vanishes close to the critical point,  that is 
as $\overline{\sigma}\rightarrow 0$. 
It is straightforward to get the term at $\omega=\mathbf{p}=0$
\begin{equation}
W(\overline{\sigma}^2)\, \overline{\sigma}^2 = 16 \int\,\frac{d^3 \mathbf{q}}{(2\pi)^3}\,
\frac{1+2 E_q^2}{ \left(E_q^2+E_q^4\right)^{3/2}} \, \overline{\sigma}^2
\end{equation}
that is
\begin{equation}
W(\overline{\sigma}^2) \simeq \frac{8}{\pi^2} \int_0^1
\frac{dq}{E_q}  \simeq - \frac{4}{\pi^2} \ln \overline{\sigma}^2\, .
\end{equation}
At $\omega=0$, by keeping the most singular contribution when $\overline{\sigma}\rightarrow 0$, we get the contribution proportional to $ \mathbf{p}^2$
\begin{equation}
Z_p(\overline{\sigma}^2) \simeq \frac{2}{\pi^2} \int_0^1
\frac{dq}{E_q}  \simeq - \frac{1}{\pi^2} \ln \overline{\sigma}^2\, .
\end{equation}
Similarly, expanding in $\omega$ ,at $ \mathbf{p}=0$, since the linear term is not diverging, the dominant contribution is proportional to $\omega^2$ and 
\begin{equation}
Z_\omega(\overline{\sigma}^2) \simeq \frac{2}{\pi^2} \int_0^1
\frac{dq}{E_q}  \simeq - \frac{1}{\pi^2} \ln \overline{\sigma}^2\, .
\end{equation}

\end{document}